# Laser induced ultrafast Gd 4f spin dynamics at the surface of amorphous Co$_x$Gd$_{100-x}$ ferrimagnetic alloys.

M. Pacé[1], D. Gupta[1], T. Ferté[1], M. Riepp[1], G. Malinowski[2], M. Hehn[2], F. Pressacco[3], M. Silly[3], F. Sirotti[4], C. Boeglin[1] and N. Bergeard[1]

[1] *Université de Strasbourg, CNRS, Institut de Physique et Chimie des Matériaux de Strasbourg, UMR 7504, F-67000 Strasbourg, France.*

[2] *Institut Jean Lamour, CNRS - Université de Lorraine, 54011 Nancy, France.*

[3] *Synchrotron SOLEIL, L'Orme des Merisiers, Saint-Aubin, 91192 Gif-sur-Yvette, France*

[4] *Physique de la Matière Condensée, Ecole Polytechnique, CNRS, 91128 Palaiseau, France*

**Abstract**:

We have investigated the laser induced ultrafast dynamics of Gd 4f spins at the surface of Co$_x$Gd$_{100-x}$ alloys by means of surface-sensitive and time-resolved dichroic resonant Auger spectroscopy. We have observed that the laser induced quenching of Gd 4f magnetic order at the surface of the Co$_x$Gd$_{100-x}$ alloys occur on a much longer time scale than that previously reported in "bulk sensitive" time-resolved experiments. In parallel, we have characterized the static structural and magnetic properties at the surface and in the bulk of these alloys by combining Physical Property Measurement System (PPMS) magnetometry with X-ray Magnetic Circular Dichroism in absorption spectroscopy (XMCD) and X-Ray Photoelectron spectroscopy (XPS). The PPMS and XMCD measurements give information regarding the composition in the bulk of the alloys. The XPS measurements show non-homogeneous composition at the surface of the alloys with a strongly increased Gd content within the first layers compared to the nominal bulk values. Such larger Gd concentration results in a reduced indirect Gd 4f spin-lattice coupling. It explains the "slower" Gd 4f demagnetization we have observed in our surface-sensitive and time-resolved measurements compared to that previously reported by "bulk-sensitive" measurements.

**Introduction**:

The discovery of deterministic helicity-independent all-optical spin switching (HI-AOS) [OST12] has driven intensive investigations on the laser induced ultrafast spin dynamics in ferrimagnetic RE-TM alloys [RAD11, LOP13, GRA13, BER14, RAD15, HIG16, HEN19]. The element- and time-resolved experiments have shown that the RE 4f and TM 3d spin dynamics occur on distinct time-scales, which is considered as the key ingredient for HI-AOS [IIH18]. They have also shown that the characteristic times associated with these ultrafast dynamical processes depend on the temperature and on the alloy composition [LOP13, FER17, CHE19, REN20, FER21, FER23]. Recently, Hennecke et al have also reported on transient magnetization in-depth gradients that appear in GdFe alloys excited by femtosecond laser pulses [HEN22]. These seminal works call for further systematic experimental and theoretical investigations to reveal the correlations between the static magnetic properties of these ferrimagnetic alloys and their unusual laser induced ultrafast spin dynamics [ATX14]. Such knowledge is of prime importance to identify the most appropriate materials for technological applications. In parallel, previous experimental works have revealed that amorphous RE-TM ferrimagnetic alloys show lateral [GRA13] and in depth non-homogeneous composition [HEB16, HAL18, INY23] as well as RE segregation at interfaces [SHE81, BER17]. On the basis of these previous observations, one can legitimately wonder whether the laser induced RE 4f spin dynamics at the surface of the alloy is affected by the composition discrepancies in respect with that in the bulk [FER23]. Disparate spin dynamics at the surface and in the bulk would be detrimental for technological applications, such as data storage devices that require thin magnetic layers [SEN21].

In this work, we have investigated the laser induced ultrafast dynamics of Gd 4f spins at the surface of $Co_xGd_{100-x}$ alloys for bulk compositions x = 80 and x = 65 by means of surface-sensitive time-resolved dichroic resonant Auger spectroscopy (TR-DRAS) [BEA13, SIL17]. We have observed that the laser induced quenching of Gd 4f magnetic order at the surface of the $Co_xGd_{100-x}$ alloys occur on a much longer time scale than that previously reported in bulk sensitive time-resolved experiments [LOP13, RAD15, FER23]. In parallel, we have characterized the structural and magnetic static properties at the surface and in the bulk of these alloys by combining Physical Property Measurement System (PPMS) magnetometry with X-ray Magnetic Circular Dichroism in absorption spectroscopy (XMCD) [NAK99] and X-Ray Photoelectron spectroscopy (XPS). The PPMS measurements confirm the nominal bulk composition of the alloys. The XPS measurements evidence the non-homogeneous composition at the surface of the alloys with a strongly increased Gd content within the first layers [SHE81,

BER17, BAL16, HAS18, INY23]. As a consequence, we infer that the larger Gd concentration in these top layers compared to that of the nominal bulk composition allows to explain the "slower" Gd 4f demagnetization we have observed in our surface-sensitive and time-resolved measurements compared to that previously obtained with "bulk-sensitive" techniques [FER23].

**Experimental**: (methods, techniques, and materials studied)

The $Co_xGd_{100-x}(20)$ alloy layers were grown by DC co-sputtering on $[Ta(5)/Cu(20)/Ta(5)]_{x5}$ multilayers deposited on Si substrates (units in nm). The layers were capped with an Al(5) layer to prevent degradation of the alloy in the ambient atmosphere [BER17]. This Al(5) layer is partly oxidized during exposure to air, but its thickness ensures that a metallic Al layer remains in contact with the $Co_xGd_{100-x}(20)$ alloys. Therefore, the actual structure of the samples is $Al_2O_3(\sim3)/Al(\sim2)/Co_xGd_{100-x}(20)/[Ta(5)/Cu(20)/Ta(5)]_5/Si$. It is worth mentioning that the samples were deposited on a 1" Si wafer which was cleaved after deposition to ensure identical samples for the various experimental techniques. In this study, we have investigated the static and dynamics magnetic properties of alloys with the nominal composition $Co_{80}Gd_{20}$ and $Co_{65}Gd_{35}$ respectively. The bulk compositions of the alloys were controlled by tuning the deposition rates of the crucibles. The actual alloy compositions were verified by recording the temperature dependence of magnetization by mean of Physical Property Measurement System (PPMS) cryostat with Vibrating Sample Magnetometer (VSM) option head (figure 1). Indeed, we have compared the temperature of magnetic compensation or the Curie temperature extracted from these measurements to the tabulated values in literature [TAO74, HAN89] to estimate the composition.

The CoGd magnetic properties were also characterized by mean of X-ray Magnetic Circular Dichroism in absorption spectroscopy (XMCD) [NAK99]. We have recorded the X-ray Absorption Spectra (XAS) at the Gd $M_{4,5}$ edges in the total electron yield acquisition mode (figure 3). These measurements were performed in the main UHV chamber on the TEMPO beamline at synchrotron SOLEIL by using circularly polarized soft X-ray [POL10]. The X-ray beam impinges the sample at an angle of $\alpha_X = 44°$ in respect with the samples normal (figure 2) since the $Co_xGd_{100-x}$ alloys display in-plane magnetic anisotropy with a preferential axis [TAY76, BER17]. The data acquisitions were performed in the remnant magnetic state after saturation along the in-plane preferential axis by a magnetic pulse of 300 ms duration and 200 Oe maximum amplitude. The XMCD spectra were obtained by recording the XAS spectra with opposite directions of the external magnetic field (figure 3a and 3c) [BER17]. The coercive field of the $Co_{65}Gd_{35}$ alloy was below 200 Oe for temperatures (T) ranging between 80 K to 300 K, while the coercive field of the $Co_{80}Gd_{20}$ alloy was below 200 Oe for T > 200K. Since we have restricted ourselves to qualitative analysis, we haven't corrected the displayed spectra for the saturation effects [NAK99]. Prior to XMCD measurements, the alloys were sputtered in the preparation chamber of the TEMPO beamline to partly remove the Al capping layer to increase the signal. We have kept a metallic Al layer to prevent the oxidization during the measurements [BER17].

Amorphous RE-TM ferrimagnetic alloys are known to display in-depth compositional gradients [HEB16, HAL18, INY23] or RE segregation at the surface [SHE81, BER17]. The composition at the surface of the alloys was thus investigated by mean of X-Ray Photoelectron Spectroscopy (XPS). The measurements were performed in the main chamber of the TEMPO beamline by using the SES2002 photoelectron analyzer. Prior to XPS measurements, the alloys were sputtered in-situ with Ar+ ions to completely remove the $Al_2O_3$(~3)/Al(~2) capping layer. However, we have regularly monitored the thickness of the remaining capping layer as a function of sputtering time in order to minimize the etching of the alloys. We have set the photon energy to 700 eV and we have recorded the XPS spectra of the Al 2p, Gd 4f, Gd 4d as well as the Co 3d and Co 3p core-levels (figure 4a and 4b). The sample's normal was aligned with the entrance of the photoelectron analyzer while the X-ray beam impinged the samples at an angle of $\alpha_X = 44°$ in respect with the sample's normal (figure 2). We have repeated these measurements at photon energies of 400 eV and 1000 eV. This protocol allows varying the surface sensitivity of core-level XPS since the electron inelastic mean free path depends on the kinetic energy of photoelectrons and thus on the photon energy (table 1) [JAB11]. As a consequence, the larger photon energies allow probing deeper into the layer. In figure 5, we show the Gd 4f and Co 3p core-levels as a function of the photon energy for the $Co_{80}Gd_{20}$ (a) and $Co_{65}Gd_{35}$ (b) alloys, respectively. A Shirley-like background was subtracted from the experimental XPS core-level peaks, while the peak's area was normalized by the photoionization cross-section of the materials [YEH85]. We have then normalized the spectra with the height of the Gd 4f peak for direct comparison. In order to quantify the Gd excess at the surface of the alloy in respect with the nominal composition, we define R as the ratio between the Gd 4f and Co 3p core-level peak's area. The dependence of R on the photon energy is depicted in figure 6a. We have developed an elementary model to provide a qualitative description of the alloy profiles [PAC23]. We have considered that the XPS signal of the specie x coming from the $i^{th}$ layer $A_x^i$ is given by $exp\left(\sum_{j=0}^{i}\left(\frac{-1}{cos(\theta).(cCo(j)*\lambda Co + cGd(j)*\lambda Gd)}\right)\right)$ with $c_x(j)$ and $\lambda_x$ the concentration of the specie x in the layers j (on top of the layer i) and the photoelectron inelastic mean free path (IMFP) respectively. The total XPS signal of specie x is then given by $A_x = \sum_{i=0}^{N} C_x^i A_x^i$ with N the alloy thickness and R is given by $\frac{A_{Gd}}{A_{Co}}$. We have used a genetic algorithm in order to determine the profile which gives the best match between the calculated dependence of R on photon energy and the experimental values (figure 6a). Considering the limited inelastic mean free path of photoelectrons in CoGd alloys (table 1), we have set the composition to the nominal value as given by VSM measurements (figure 1) for thicknesses above 5 nm.

The laser induced ultrafast Gd 4f spin dynamics was recorded by using TR-DRAS [BEA13, SIL17] at the TEMPO Beamline of synchrotron SOLEIL [POL10] by using both the hybrid (figure 7) and low-alpha (figure 8) filling mode [SIL17]. This technique offers surface sensitivity, element specificity, the sensitivity to 4f magnetic order and time-resolutions of 60 ps or 12 ps in the hybrid and low-alpha filling mode, respectively. In our experiments, the energy of the circularly polarized X-ray pulses matches the Gd $M_5$ absorption edge ($E_{hv}$=1192.7eV) while the Auger photoelectrons ($E_{KE}$=1184.5eV) are collected by a Scienta 2002 photoelectron analyzer equipped with a delay line [BER11]. The kinetic energies of Auger photoelectrons allow separating them from the photoelectrons generated by laser absorption [SIR14]. The intensity of the generated Auger photoelectrons (in e-/sec) is proportional to the X-ray absorption in the $Co_xGd_{100-x}$ layers. Since the X-ray absorption of circularly polarized X-ray in magnetic materials depends on the magnetization, the Auger photoelectron yield depends on the magnetization, as depicted in figures 7(a,b) and figures 8(a,b). The difference between the Auger photoelectrons collected for two opposite directions of the magnetic field (H+ and H-) is proportional to the magnetization [SIL17]. The variation in the amplitude of the signal proportional to the magnetization is indicated by the blue M arrows in the figures. The TR-DRAS experiments have consisted in recording the Auger photoelectrons distribution for both magnetic field helicities as a function of the delay between the IR pump and the X-ray probe pulses by using either the hybrid (time resolution ~60 ps) or the low alpha (time resolution ~12ps) modes. In figures 7(a,b) and 8(a, b), we display the Auger photoelectron yield for two magnetic field helicities at negative delay (a) and t=100 ps after laser excitation (b) for hybrid and low-alpha modes, respectively. Then, the photoelectron Auger yields are integrated over the investigated kinetic energy range. The difference of the integrated signal between both magnetic field helicities as a function of the pump-probe delay (figure 7c and 8c) is labelled TR-DRAS in the following. The laser frequency was set to 141 kHz, which is six times lower than the frequency of the isolated electron bunch in the SOLEIL filling pattern. The thick

Ta/Cu/Ta buffer layer beneath the $Co_xGd_{100-x}$ alloys allows to enhance the heat dissipation during the pump-probe experiments, ensuring moderate temperature elevation by DC-heating. The laser beam impinges the sample at an angle of 67° in respect with the sample's normal (figure 2). For each delay step, a pulsed magnetic field of 300 ms duration, 200 Oe amplitude is applied along the magnetic easy axis to saturate the alloys. The acquisition is then performed at remanence ($H_{ext}$ = 0 Oe), which prevents perturbation of the photoelectron trajectories towards the analyzer. The detection scheme allows collecting separately the photoelectron pulses generated by every isolated X-ray pulses, which means that between two successive laser excitations, 1 X-ray pulse probes the excited magnetic state of the alloy while the next 5 pulses probe the non-excited magnetic state of the alloy. Thus, we have direct evidences that the samples are not damaged during the acquisition and that the magnetization of the alloy is restored to its equilibrium state between two successive excitations even in absence of external magnetic field. It also allows normalizing the photoelectron Auger yield coming from the excited magnetic state by the photoelectron Auger yield coming from the non-excited magnetic state to increase significantly the signal to noise ratio [SIL17]. The time-resolved experiments were carried out for the $Co_{65}Gd_{35}$ and $Co_{80}Gd_{20}$ alloys (figures 7 and 8). We have not fully removed the Al protective layer during the time-consuming pump-probe experiments to avoid oxidization of the top layers [BER17]. The cryostat temperatures were set to 80 and 200 K while the laser fluences were 2.1 and 2.8 mJ/cm² respectively. The higher temperature for the $Co_{80}Gd_{20}$ alloy was needed to reach a coercive field below 200 Oe, while the larger laser fluence was selected to obtain similar demagnetization amplitudes. We have restricted the laser fluence to 2.8 mJ/cm² because we observed traces of degradation at the surface of the alloys for laser fluence above 3 mJ/cm². Furthermore, we have restricted our investigations to moderate excitation (~50 % demagnetization) for direct comparison with previous bulk-sensitive XMCD experiments performed in pure Gd [WIE11, ESC14] and CoGd alloys [FER23]. As a consequence, for the $Co_{65}Gd_{35}$ ($Co_{80}Gd_{20}$) alloy, the equilibrium temperature was below (above) the temperature of magnetic compensation [FER17]. The laser induced dynamics of Gd 4f spins in the $Co_{65}Gd_{35}$ alloy was also recorded on a broader delay range by using the hybrid mode (figure 7). Such measurements allow to determine accurately the characteristic recovery times in order to lower the uncertainties in fitting the TR-DRAS data obtained in the low-alpha mode. For the $Co_{80}Gd_{20}$ alloy, the delay range investigated in the low-alpha operation mode was sufficient to estimate the characteristic recovery times.

**Experimental results and discussion**:

The net magnetization of the alloys as a function of temperature measured by PPMS-VSM magnetometry and XMCD spectroscopy are plotted in figure 1 and figure 3 respectively. We observe that the magnetization of the $Co_{65}Gd_{35}$ alloy goes to zero at T ~ 370 K (figure 1), as expected from tabulated data [TAO74, HAN89]. We also observe that the magnetization of the $Co_{80}Gd_{20}$ alloy crosses zero at T = 170 K (figure 1), which is the temperature of magnetic compensation ($T_{comp}$). This value for $T_{comp}$ is also consistent with the nominal composition as attested by tabulated data [HAN89]. The opposite signs of the XMCD signal at the Gd $M_5$ edges for the $Co_{80}Gd_{20}$ alloy (figure 3b) and the $Co_{65}Gd_{35}$ alloy (figure 3d) confirm that the XMCD spectra were acquired above (below) the temperature of magnetic compensation for the $Co_{80}Gd_{20}$ ($Co_{65}Gd_{35}$) alloy. The characterization of the magnetic properties by mean of PPMS-VSM magnetometry and XMCD spectroscopy show that the actual composition of the alloy in the bulk matches the nominal composition.

After ensuring that the alloys have the correct composition in the bulk, we focused on the electronic and structural properties at the surface of the alloy by means of XPS (figure 4). In figure 5a and 5b, we observe that the height of the Co 3p peak increases in respect with that of the Gd 4f peak when the photon energy increases for both samples. It demonstrates that the surface of the alloys is less concentrated in Co than the bulk. Our model gives a qualitative agreement with the experimental variation of R (figure 6a). The profile which corresponds to the calculated variation of R with photon energy is depicted in figure 6b. In the case of $Co_{80}Gd_{20}$ alloy, the simulated profile shows a segregated Gd layer on top of an almost homogeneous sample with almost the nominal composition as previously reported for a comparable alloy composition [BER17]. However, the simulated profile of the $Co_{65}Gd_{35}$ alloy shows a pronounced Gd compositional gradient at the surface of the alloy and thus, a sizable increase of the Gd contents over almost 2 nm in respect with the nominal composition. A more accurate determination of the profile given for layers deeper than 2 nm should be confirmed by other experimental techniques with higher probing depth. Our observations based on XPS demonstrate that the alloys are non-homogeneous and that these non-homogeneity's discrepancies depend on the alloy composition.

The TR-DRAS measurements show a quenching of the Gd 4f magnetic order which is followed by a recovery for both alloys (Figure 7c, 8c). However, our data show different dynamical responses to laser excitation for the Gd sublattices in both alloys. Indeed, we observe that the maximum demagnetization amplitudes are reached at delays $d_{max}$ = 28 ps and $d_{max}$ = 100 ps for the $Co_{80}Gd_{20}$ and the $Co_{65}Gd_{35}$ alloys respectively (figure 8). The data were adjusted by exponential functions convoluted with a Gaussian function to extract the characteristic demagnetization ($\tau_M$) and recovery ($\tau_R$) times (solid lines in figures 7c and 8c) [BOE10, LOP12, BER14]. We have estimated $\tau_M$ = 12 ps and $\tau_R$ = 55±10 ps for the $Co_{80}Gd_{20}$ alloy and $\tau_M$ = 37±10 ps and $\tau_R$ = 400±60 ps for the $Co_{65}Gd_{35}$ alloy. For the $Co_{80}Gd_{20}$ alloy, we have set a lower limit of 12 ps for $\tau_M$ during the fit processing which is given by the experimental time resolution. For the $Co_{65}Gd_{35}$ alloy, the recovery time ($\tau_R$ = 400±60 ps) was extracted from the data acquired in the hybrid mode (figure 7c).

The faster recovery for the $Co_{80}Gd_{20}$ alloy ($\tau_R$ = 55±10 ps) measured at T = 200K compared to the $Co_{65}Gd_{35}$ alloy ($\tau_R$ = 400±60 ps) measured at T = 80K is probably partly caused by the temperature dependent thermal conductivity in RE-TM alloys as shown by Hopkins et al. in FeCoGd alloys [HOP12]. Such temperature dependent recovery for Gd 4f spins in CoGd alloys has been reported recently [FER23]. Furthermore, the article published by Ferté et al. allows explaining that the characteristic demagnetization time $\tau_M$ < 12 ps for the $Co_{80}Gd_{20}$ alloy is shorter than that for the $Co_{65}Gd_{35}$ alloy ($\tau_M$ = 37±10 ps), because of the different Gd composition of the alloys. Indeed, in CoGd alloys, the indirect Gd 4f spin-lattice coupling is enhanced in respect with pure Gd layers by the Gd 4f – Gd 5d intra-atomic exchange coupling, the Co 3d – Gd 5d inter-atomic exchange coupling and the Co 3d spin-lattice coupling [REN19, HEN19, FER23]. By reducing the Co concentration at the surface of the alloy, the indirect Gd 4f spin-lattice coupling is also reduced. However, the laser induced dynamics of Gd 4f spins on top of the $Co_{65}Gd_{35}$ alloy shows similarities with that reported by Wietstruk et al. for pure Gd layer in low-alpha mode at BESSY [WIE11]. Indeed, although the time-resolution of our TR-DRAS experiment and the statistic do not allow for resolving the two-step demagnetization, it is worth noticing that the demagnetization times we extracted from our experimental data ($\tau_M$ = 37±10 ps) is similar to the characteristic times of the "slow" demagnetization process of pure Gd [WIE11] rather than that reported by means of bulk-sensitive TR-XMCD in $Co_{72}Gd_{28}$ alloy ($\tau_M$ ~ 4 ps [FER23]). This observation is thus qualitatively consistent with the rich Gd content at the surface of the $Co_{65}Gd_{35}$ alloys, which induces disparate Gd 4f spin dynamics compared to that expected within the bulk. It is worth noticing that Graves et al. have shown that lateral composition gradient also influence the spin dynamics in RE-TM alloys [GRA13]. However, the amplitude in composition non-homogeneities they have reported are much lower than that we estimate at the surface of our layers. In the case of the $Co_{80}Gd_{20}$ alloy, any discussion regarding the characteristic demagnetization time for the Gd 4f sublattice and its comparison with previous bulk-sensitive measurements [LOP13, RAD15] are hampered by the time-resolution of our experiment. Our works show that an accurate description of the Gd 4f spin dynamics in ferrimagnetic alloys requires particular attention to their structural properties. Our work also calls for further in-depth resolved experiments in RE-TM alloys with nanometer spatial resolution combined with sub-picosecond time resolutions either at large scale facilities [JAL17] or by using table-top X-ray sources [HEN22].

**Conclusions:**

In this work, we propose an experimental investigation on the structural and static magnetic properties as well as on the laser induced ultrafast Gd 4f spin dynamics at the surface of $Co_xGd_{100-x}$ ferrimagnetic alloys. We have shown that the surface of the CoGd alloys is Gd-rich compared to the bulk and these non-homogeneities are composition dependent. Indeed, the Gd migration at the surface is more pronounced in alloys that display a larger Gd content in the bulk. We have used surface sensitive time-resolved dichroic resonant Auger spectroscopy to show that the Gd 4f spin dynamics is much slower at the surface of $Co_{65}Gd_{35}$ alloy compared to that previously reported in the bulk of comparable alloys by means of time-resolved XMCD measurements. This discrepancy can be explained by the larger Gd composition on the top of the alloys which results in a reduced indirect Gd 4f spin-lattice coupling. However, our preliminary work gives a partial view of the Gd 4f spin dynamics because of the limited time-resolution and calls for further in-depth and time-resolved experiments. Among the point to be clarified, we hint that the exchange coupling between the Gd-rich surface and the bulk of the alloy may also play a role on the Gd 4f spin dynamics by modifying the effective Curie temperature ($T_{Curie}$). $T_{Curie}$ is known to be a key parameter to determine the characteristic demagnetization times in the framework of the microscopic 3 temperatures model [KOO10]. These issues have to be addressed for future technological applications which require a fine tuning of the characteristic demagnetization times [REM20] or a downsizing of the ferrimagnetic alloy layers [LAL17, SEN21].

**Acknowledgements:**

We acknowledge SOLEIL for provision of synchrotron radiation facilities in using the beamline TEMPO. This project has received fundings from the European Union's Horizon 2020 research and innovation program under the Marie Skłodowska-Curie grant agreement number 847471, the French national agency for research ANR-20-CE42-0012-01 and from the Région Grand Est. The competence center Magnetism and Cryogenics, Institut Jean Lamour, is acknowledged for magnetometry measurements.

**Table 1:** Mean free paths of the different core-level photoelectrons Co3d, Co3p, Gd4f and Gd 4d as a function of the incident photon energy as given by [JAB11].

| Photon energy (eV) | Mean free path λ (Å) | | | |
|---|---|---|---|---|
| | λCo$_{3d}$ | λCo$_{3p}$ | λGd$_{4f}$ | λGd$_{4d}$ |
| **400** | **6.6** | **6.2** | **9.8** | **7.3** |
| 700 | 8.5 | 8.2 | 14.7 | 12.5 |
| 1000 | 10 | 9.7 | 19.4 | 17.3 |

**Figures:**

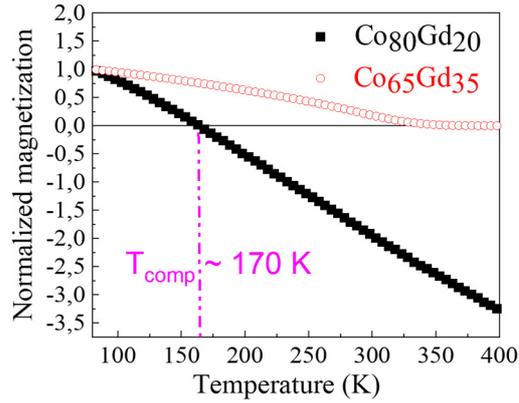

*Figure 1: Magnetization as a function of temperature measured by mean of VSM magnetometry for the $Co_{80}Gd_{20}$ (black filled squares) and $Co_{65}Gd_{35}$ (red empty circles) alloys. The values are normalized by the magnetization at T = 80K.*

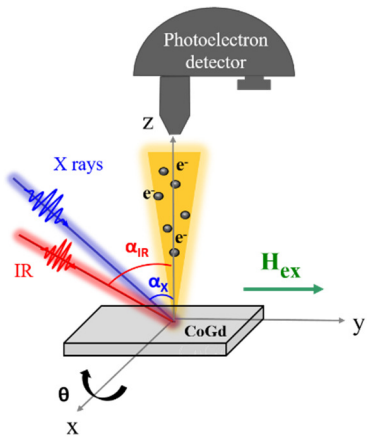

*Figure 2: Geometry of the experiment where the external magnetic field can be applied in the plane of the films. Incidence angles of the X rays and IR pump are defined by $α_X$ (=44°) and $α_{IR}$ (=67°) in respect with the surface normal.*

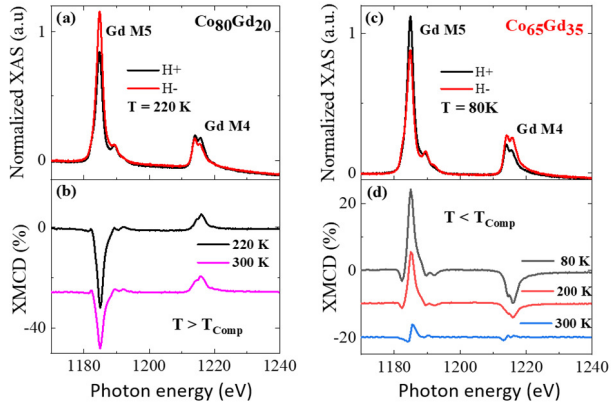

*Figure 3: X-ray absorption spectra for two opposite helicities of the external magnetic field (black and red curves) at the Gd $M_{4,5}$ absorption edges for the $Co_{80}Gd_{20}$ alloy at T = 220K (a) and for the $Co_{65}Gd_{35}$ alloy at T = 80 K (c). (b) The XMCD spectra at the Gd $M_{4,5}$ absorption edges for the $Co_{80}Gd_{20}$ alloy at T = 220K and T = 300K. (d) The XMCD spectra at the Gd $M_{4,5}$ absorption edges for the $Co_{65}Gd_{35}$ alloy at T = 80K, T = 200K and T = 300K.*

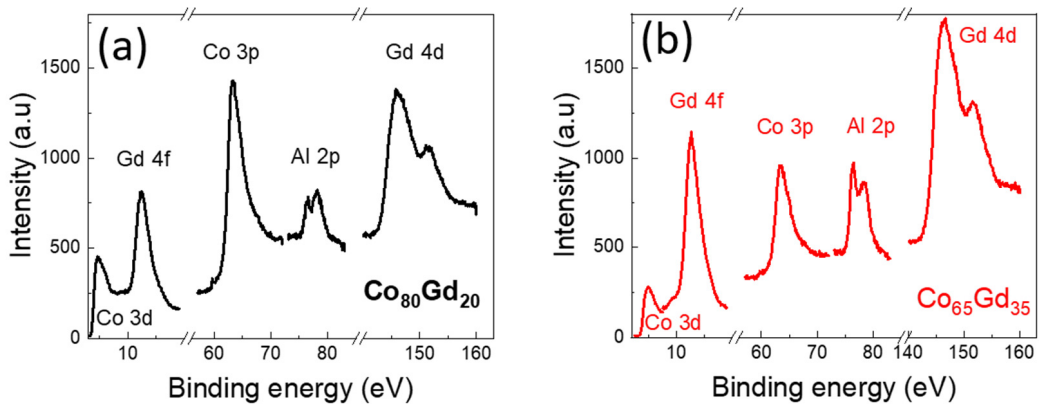

*Figure 4: X-ray photoelectron spectra taken at 700 eV photon energy showing the Co 3p, 3d, Gd 4f, 4d and the Al 2p core-levels for the alloys (a) $Co_{80}Gd_{20}$ and (b) $Co_{65}Gd_{35}$.*

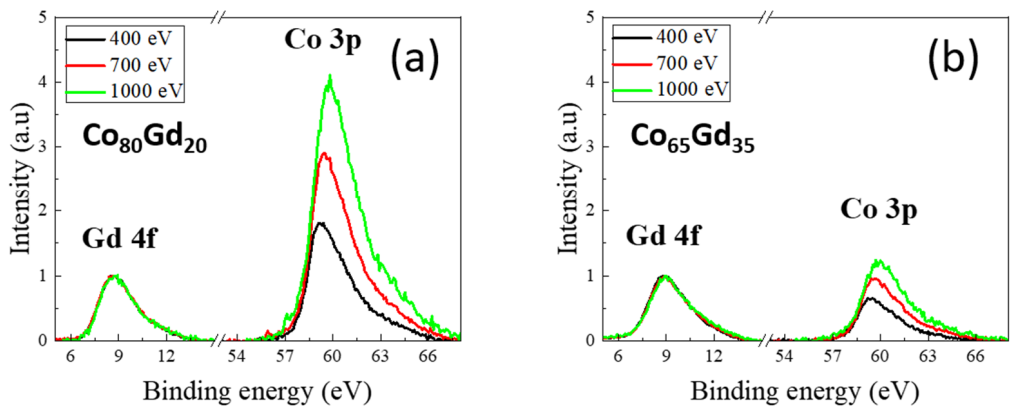

*Figure 5: X-ray Photoelectron spectra of the Co 3p and Gd 4f core-levels for (a) $Co_{80}Gd_{20}$ and (b) $Co_{65}Gd_{35}$ at the photon energies of 400 eV (black spectra), 700 eV (red spectra) and 1000 eV (green spectra). The spectra have been subtracted by a*

Shirley function and normalized by the photoionization cross section [JAB11]. All spectra have been normalized to 1 at the Gd4f.

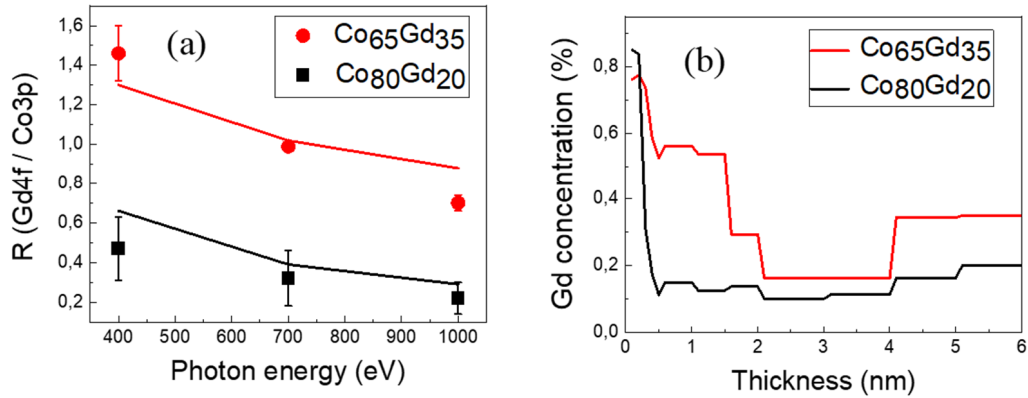

Figure 6: (a) Comparison between the experimental (symbols) and the calculated (solid lines) energy dependence of the ratio R between the Gd 4f and Co 3p core levels. (b) Simulated Gd composition profile at the surface of the $Co_{80}Gd_{20}$ (black line) and the $Co_{65}Gd_{35}$ (red line) alloys

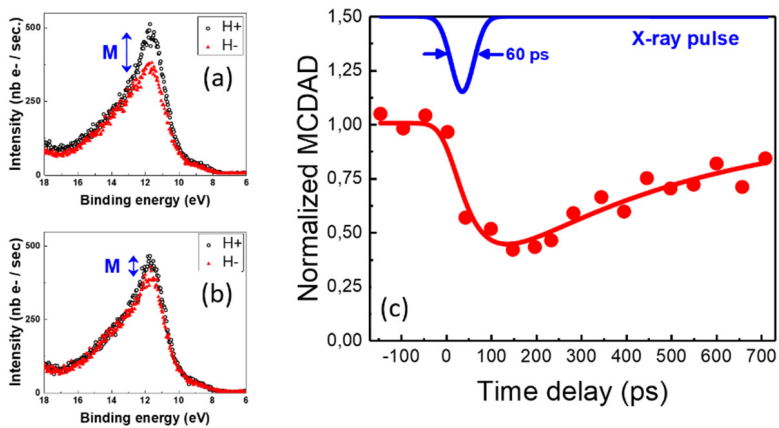

Figure 7: Auger electron yield for two magnetic field helicities at negative (a) and positive (b) delays for the $Co_{65}Gd_{35}$ alloy measured by using the hybrid filling mode (time resolution 60 ps). (c) Normalized magnetic circular dichroism as a function of the pump-probe delay for the $Co_{65}Gd_{35}$ alloy (red filled circle) at T = 80 K. The solid lines are exponential fit as described in text.

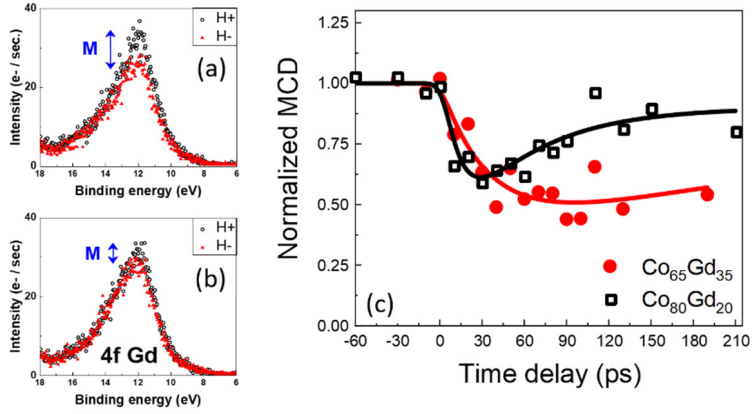

*Figure 8: Auger electron yield for two magnetic field helicities at negative (a) and positive (b) delays for the $Co_{65}Gd_{35}$ alloy measured by using the low-α filling mode (time-resolution 12 ps). (c) Normalized magnetic circular dichroism as a function of the pump-probe delay for the $Co_{65}Gd_{35}$ (red filled circle) at T = 80 K and the $Co_{80}Gd_{20}$ at T = 200 K (black empty squares) alloys. The solid lines are exponential fit as described in text.*